\documentclass[conference]{IEEEtran}

\usepackage{graphicx}
\usepackage{subcaption}   
\usepackage{algorithm}
\usepackage{algpseudocode}
\usepackage{enumitem}

\usepackage{amsmath}

\usepackage{mathtools}   
\AtBeginDocument{%
  }

\newboolean{showcomments}
\setboolean{showcomments}{true}
\makeatletter
\newcommand{\mynote}[3]{%
  \ifthenelse{\boolean{showcomments}}{%
   \fbox{\bfseries\sffamily\scriptsize#1}%
   {\small$\blacktriangleright$\textsf{\emph{\color{#3}{#2}}}$\blacktriangleleft$}}%
  {%
   \@bsphack
   \@esphack
  }%
}
\makeatother

\usepackage[printonlyused]{acronym}
\acrodef{ALCF}[ALCF]{Argonne Leadership Computing Facility}

\newcommand{\ke}[1]{\textcolor{black}{#1}}

\usepackage{tikz}

\usepackage{pifont}

\usepackage{xspace}

\newcommand{\tool}{\textsc{{CHiArA}}\xspace}

\usepackage{cleveref}
\usepackage{url}
\crefformat{section}{\S#2#1#3} 
\crefformat{subsection}{\S#2#1#3}
\crefformat{subsubsection}{\S#2#1#3}

\usepackage{todonotes}

\begin{document}

\title{Configurable and Hierarchical Allreduce}






\author{
\IEEEauthorblockN{
Valentino Guerrini\IEEEauthorrefmark{1}\IEEEauthorrefmark{3},
Ke Fan\IEEEauthorrefmark{2},
Sidharth Kumar\IEEEauthorrefmark{1}
}

\IEEEauthorblockA{
\IEEEauthorrefmark{1}
University of Illinois Chicago,
Chicago, IL, USA\\
Email: \{vguer24, sidharth\}@uic.edu
}

\IEEEauthorblockA{
\IEEEauthorrefmark{2}
Temple University,
Philadelphia, PA, USA\\
Email: ke.fan@temple.edu
}

\IEEEauthorblockA{
\IEEEauthorrefmark{3}
Politecnico di Milano,
Milan, Italy
}
}


\maketitle

\begin{abstract}
\texttt{MPI\_Allreduce} is among the most performance-critical collectives in
large-scale scientific computing and distributed machine learning, yet the
small- and medium-message regime remains challenging: latency, synchronization
depth, and strong hardware hierarchy between intra- and inter-domain communication
all compound per-invocation cost.
We present \tool, a configurable hierarchical Allreduce that encodes hardware
hierarchy through a logical batch--lane topology and executes a staged schedule
in which only a bounded portion of the reduction vector is active at a time.
Inter-batch communication is distributed across multiple ranks via a rotating-root
lane primitive, avoiding centralized leaders. \tool further enables a
\emph{semi-composed} Rabenseifner-style Allreduce by preserving a lane-aligned
intermediate layout across the Reduce-Scatter/Allgather boundary, eliminating
redundant intra-domain reorganization.
We evaluate \tool on Polaris, Aurora, and Fugaku, achieving speedups of up to
$1.94\times$, $13.43\times$, and $13.48\times$ over vendor \texttt{MPI\_Allreduce},
and up to $2.2\times$ end-to-end speedup in a parallel $k$-means application.
\end{abstract}





    \section{INTRODUCTION}


Collective communication is a critical component of large-scale scientific
computations, data-intensive workflows, and distributed machine learning.
Operations such as \emph{Allreduce}, \emph{Allgather}, and \emph{Reduce-Scatter}
are invoked repeatedly to exchange intermediate values, synchronize algorithmic
state, and compute global reductions. Representative applications include Krylov
subspace solvers~\cite{Yamazaki2017,Lockhart2022}, multidimensional FFT-based
simulations~\cite{Ayala2021}, ensemble and multi-trajectory workflows~\cite{Wozniak2019},
and distributed deep learning systems~\cite{Akiba2017}. These workloads may issue
tens of thousands of collectives during a single execution, so even modest
inefficiencies can accumulate into substantial runtime overhead at scale.

Achieving high performance in the \emph{small- and medium-message regime}
(roughly 64~B to 32~KB per rank in our experiments) remains particularly
challenging. Unlike large messages where bandwidth dominates, small messages are
sensitive to both the latency term $\alpha$ and the per-round overheads implied
by the $\alpha$--$\beta$ model\cite{hockneymodel}, including synchronization and message-matching
costs. Modern HPC platforms~\cite{sato2021co, bertoni2024early} amplify this challenge by exposing strong
communication hierarchy: exchanges within a fast shared-memory domain (e.g.,
within a socket or node) are fundamentally cheaper than exchanges that traverse
the network fabric (e.g., Slingshot, InfiniBand, or Tofu-D). As a result, reducing
global synchronization depth and restricting expensive communication to a
structured subset of the algorithm are central to improving performance in this
regime.

\begin{figure}[t]
    \centering
    \includegraphics[width=\columnwidth]{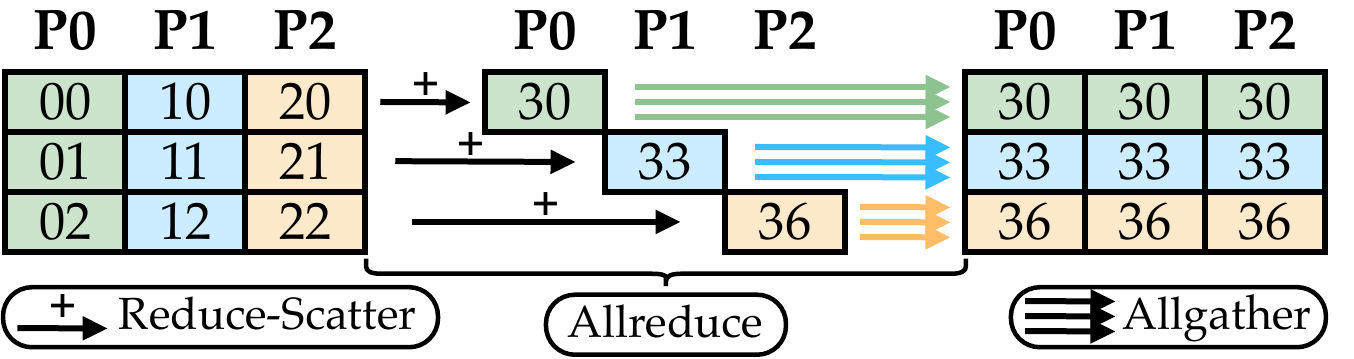}
    \caption{Rabenseifner-style Allreduce decomposition into Reduce-Scatter and Allgather~\cite{Rabenseifner2004}.}
    \label{fig:intro}
    \vspace{-0.5cm}
\end{figure}

State-of-the-art implementations of Allreduce are typically built on
classical tree- and ring-based algorithms, as well as composite schemes
that trade latency for bandwidth across message-size regimes
\cite{PatarasukYuan2009}. Modern MPI libraries already incorporate
hierarchical designs that distinguish between intra-node and inter-node
communication to exploit shared memory within nodes. However, these
hierarchies are often implicit, fixed by the library implementation,
and not directly exposed to algorithm designers or users.

Building on this observation, prior work has proposed node- and
topology-aware collectives that optimize these hierarchical
strategies by improving aggregation, scheduling, and topology
mapping \cite{BienzNodeAware2019,NetLoc2014}. While these approaches
demonstrate substantial performance benefits, they typically rely on
predefined hierarchical decompositions and focus on optimizing specific
levels of the hierarchy, offering limited flexibility to adapt the
overall communication structure to varying node sizes, socket counts,
or system scales.

More recent research has explored generalized collective frameworks that expose 
algorithmic parameters, such as radix, to better exploit modern hardware features. 
Wilkins et al. propose system-agnostic generalized communication kernels with tunable 
radix values, demonstrating that variable-radix designs can outperform fixed-structure 
implementations across multiple platforms \cite{exascale}. 
While effective at improving generality and adaptability, this work does not 
explicitly target hierarchical decomposition within nodes or coordinated tuning 
across Allreduce phases.

Overall, existing Allreduce implementations address different parts of
the performance space, but with complementary limitations. Latency- and
bandwidth-optimized algorithms rely on largely fixed communication
structures, while topology-aware approaches operate at
the granularity of whole collectives rather than exposing control over
their hierarchical organization. On modern HPC systems with strong
asymmetry between intra- and inter-node communication, these limitations
restrict the ability to systematically balance latency, bandwidth, and
hierarchy. This gap motivates Allreduce designs that expose a small set
of explicit parameters and directly encode hierarchical structure.

This work is driven by the observation that, on modern hierarchical HPC systems,
the performance of \texttt{MPI\_Allreduce} in the small- and medium-message regime
is primarily limited by synchronization depth and redundant data movement across
locality boundaries rather than by peak bandwidth. While bandwidth-efficient
designs based on Reduce-Scatter followed by Allgather are widely used, their
hierarchical realizations typically introduce additional intra-node
reorganization steps between phases and often centralize cross-domain communication
on a small subset of ranks, creating injection and progress bottlenecks.

To address this gap, we develop \tool, a \underline{C}onfigurable \underline{Hi}erarchical \underline{A}ll\underline{r}educe \underline{A}lgorithm inspired by the Reduce-Scatter + Allgather factorization popularized by Rabenseifner~\cite{Rabenseifner2004} (Figure~\ref{fig:intro}).
CHiArA is built on three insights. First, expressing hierarchy through a logical
batch--lane topology exposes locality while allowing inter-batch communication to
be distributed across multiple ranks instead of a single leader. Second, staging
the reduction vector bounds the working set and decouples the inter-batch schedule
from the full message size. Third, preserving a lane-aligned intermediate layout
across Reduce-Scatter and Allgather eliminates redundant intra-batch reorganization,
enabling a semi-composed Allreduce that reduces synchronization and local data
movement without altering the inter-batch communication pattern. This paper makes the following contributions:

\begin{enumerate}[leftmargin=*]
  \item We present \tool, a hierarchical Allreduce with a configurable batch--lane
  topology that confines the majority of communication to fast shared-memory and
  exposes a set of tuning parameters, including batch size and independent
  intra-batch radices for Reduce-Scatter and Allgather.
  \item We introduce a staged Reduce-Scatter and Allgather design that preserves a
  lane-aligned intermediate layout, enabling a \emph{semi-composed} Allreduce that
  eliminates redundant intra-batch reorganization otherwise incurred when composing
  standalone hierarchical collectives.
  \item We evaluate CHiArA on three leadership-class systems (Polaris, Aurora, and
  Fugaku), demonstrating speedups of up to $1.94\times$, $13.43\times$, and
  $13.48\times$, respectively, over vendor \texttt{MPI\_Allreduce}.
  \item We integrate CHiArA as a drop-in replacement into an open-source MPI+OpenMP
  $k$-means implementation, achieving end-to-end iteration speedups.
\end{enumerate}

\section{Background}
\label{sec:background}

This section reviews the collective communication algorithms and
patterns that form the basis of the hierarchical and parameterized
design developed in this paper. We focus on the algorithmic
building blocks used for Allgather, Reduce-Scatter, and Allreduce.

\textit{(a) Recursive doubling and recursive halving:}
Recursive doubling is a classic logarithmic collective algorithm with
radix~2, commonly used for Allgather and Allreduce in latency-dominated
regimes. In each of $\log_2 P$ rounds, a process exchanges data with a
single partner whose rank differs by a power of two. Recursive halving
is the reduction analogue and forms the basis of many Reduce-Scatter
implementations. These algorithms are primarily used for small messages, 
where latency dominates. By completing in $O(\log P)$ rounds, recursive doubling minimizes synchronization overhead 
and outperforms linear schemes in latency-dominated regimes\cite{hammer, thakur_2005}.
For Allgather, the exchanged payload grows across rounds, whereas 
for Reduce-Scatter via recursive halving the payload
shrinks as the vector is progressively partitioned.

\textit{(b) Ring-based algorithms:}
Ring algorithms arrange processes in a logical cycle and perform
nearest-neighbor communication over $P-1$ rounds. Each process sends and
receives a fixed-size data chunk per round, enabling efficient pipelining and
high bandwidth utilization. Ring-based collectives are widely used in production MPI
libraries and form the core of bandwidth-optimal Allreduce implementations for
large messages, where minimizing total data volume dominates performance.

\textit{(c) Rabenseifner’s Allreduce:}
A widely adopted approach for implementing Allreduce is Rabenseifner’s
method\cite{Rabenseifner2004}, which decomposes the operation into a \texttt{Reduce-Scatter} followed by an \texttt{Allgather} (two phases). This factorization reduces redundant communication and
allows combining bandwidth-efficient algorithms for the two phases.
Rabenseifner style Allreduce is used in MPICH, OpenMPI, and vendor MPI
implementations and provides the structural foundation for the Allreduce
design studied in this work.


\textit{(d) Recursive exchange and radix-$k$ algorithms:}
Recursive exchange generalizes recursive doubling by allowing an
arbitrary radix $2 \le k \le P-1$. Each round communicates with $k-1$
peers, reducing the number of rounds to $\lceil \log_k P \rceil$ at the
cost of increased per-round data movement. This defines a continuous design space between logarithmic-depth
algorithms and high-fanout, low-depth exchanges. Radix-$k$ exchange
patterns form the basis of several generalized Allgather,
Reduce-Scatter, and Allreduce schemes proposed in the literature\cite{exascale}.

\textit{(e) Hierarchical collectives:}
Modern MPI implementations exploit node-level hierarchy by separating
intra-node and inter-node communication. Collective operations are
decomposed into local shared-memory phases and global network phases,
reducing inter-node traffic and improving locality. However, this hierarchical 
decomposition is typically fixed at the
physical node granularity and implemented using a small set of
predefined algorithms (e.g., master-based intra-node reduction)\cite{BienzNodeAware2019},
with limited exposure of algorithmic parameters for tuning or
adaptation.

\section{Design of \tool }
\label{sec:design-new}

\tool revisits the classical Rabenseifner \texttt{Allreduce} under modern HPC hierarchical systems. While Rabenseifner achieves communication efficiency through a two-phase decomposition consisting of \texttt{Reduce-Scatter} followed by \texttt{Allgather}, practical implementations (e.g., in MPICH~\cite{MPICH} and OpenMPI~\cite{OpenMPI}) incorporate hierarchy using fixed intra- and inter-node data exchanges. However, such designs rely on fixed decomposition boundaries and treat the two phases independently, resulting in excessive synchronization depth, redundant data movement at the phase boundary, and imbalanced inter-node communication (see Section~\ref{sec:background} (c) for details).

\tool addresses these limitations by jointly introducing \emph{explicit} and \emph{tunable} hierarchy. The design overview of \tool is detailed in Section~\ref{sec:design-new:overview}. Specifically, \tool designs an efficient \emph{batch--lane} structure that exposes hardware hierarchy while distributing global communication across ranks to maximize network utilization (Section~\ref{sec:design-new:topology}). It further introduces a staged execution model that decomposes data into smaller blocks, limiting the working set and balancing workload across ranks (Section~\ref{sec:design-new:decomp}). Finally, \tool enables a small set of tunable parameters that directly control the hierarchy granularity and the communication schedules to adapt to various hardware and workloads (Section~\ref{sec:design-new:tunable}).



\begin{table}[t]
\centering
\footnotesize
\renewcommand{\arraystretch}{1.15}
\caption{Definitions of Notations}
\label{tab:notation}
\begin{tabular}{|c|p{0.4\columnwidth}|c|}
\hline
$P$ & Total number of ranks (processes) & $\ge 1$ \\
\hline
$PPN$ & Number of ranks per node & $0 \le PPN < P$ \\
\hline
$b$ & Batch size (ranks per batch) & $1 \le b \le b_{\max}$ \\
\hline
$B$ & Number of batches ($B=P/b$) & $B \ge 1$ \\
\hline
$x$ & Batch id of rank $r$ ($x=\lfloor r/b\rfloor$) & $0 \le x < B$ \\
\hline
$i$ & Lane id of rank $r$ ($i=r\bmod b$) & $0 \le i < b$ \\
\hline
$m$ & Elements contributed per rank & $\ge 0$ \\
\hline
$X_d$ & Block $d$ of the input vector (size $m/B$) & $0 \le d < B$ \\
\hline
$I$ & Number of stages ($I=\lceil B/b\rceil=\lceil P/b^2\rceil$) & $\ge 1$ \\
\hline
$\rho$ & Root batch for stage $t$, lane $i$ ($\rho=tb+i$) & $0 \le \rho$; active if $\rho<B$ \\
\hline
$k_{\mathrm{RS}}, k_{\mathrm{AG}}$ &Radices for Reduce-Scatter and Allgather & $2 \le k_{\mathrm{RS}}, k_{\mathrm{AG}} \le b-1$ \\
\hline
\end{tabular}
\end{table}

\subsection{Design Overview}
\label{sec:design-new:overview}


The core design principle of \tool is to enable hierarchical, fully distributed communication with explicit tunable control, eliminating centralized bottlenecks while adapting to hardware and workload diversity. At a high level, the algorithm decomposes into four structured phases: local reduce-scatter, global reduction, global broadcast, and local allgather. The first two phases together realize a hierarchical Reduce-Scatter, while the latter two implement a hierarchical Allgather.
Local reduce-scatter first combines data within small local groups, reducing the data volume per participant in global communication and exposing parallelism. Global reduction then aggregates these partial results across groups, distributing communication across multiple participants to avoid centralized bottlenecks. Global broadcast subsequently disseminates the reduced results back to all groups without introducing additional coordination overhead. Finally, local allgather reconstructs the full result within each group, leveraging efficient local communication.

With this design, ranks are grouped into \emph{batches} that serve as locality domains and map communication onto fast intra-node bandwidth. The batch size is tunable from 1 to the number of ranks per node (\emph{PPN}), allowing flexible control over the locality granularity. This grouping separates communication into intra-batch and inter-batch phases, enabling locality-aware execution. However, routing inter-batch communication through a single representative per batch leads to serialization and underutilization of network bandwidth.
To address this limitation, \tool introduces \emph{lanes}: ranks with the same intra-batch index across batches form independent communication paths, allowing multiple ranks to participate in inter-batch reduction and broadcast concurrently. This design distributes communication load across participants and eliminates centralized bottlenecks.
Together, batches and lanes form the essential \emph{batch--lane} structure of \tool.

Importantly, this structure aligns naturally with the Reduce-scatter/Allgather boundary. Lane-parallel inter-batch reduction produces disjoint sets of fully reduced blocks, and inter-batch broadcast reuses the same lanes to redistribute them. As a result, no additional data reorganization is required between phases, and both proceed with the same distributed execution pattern without introducing extra synchronization overhead.

To map computation onto this structure, the input vector per rank is uniformly partitioned into \emph{blocks}, one per batch, ensuring balanced workload distribution. Execution proceeds in \emph{stages}, each activating a subset of lanes to process their assigned blocks. This staged execution bounds the working set, enables pipelined progress, and reduces global synchronization depth. We now formalize these structural concepts used in \tool.

\subsection{Logical batch–lane Topology}
\label{sec:design-new:topology}

\begin{figure}[t]
  \centering
  \includegraphics[width=\columnwidth, trim=0 2.5cm 0 2.5cm, clip]{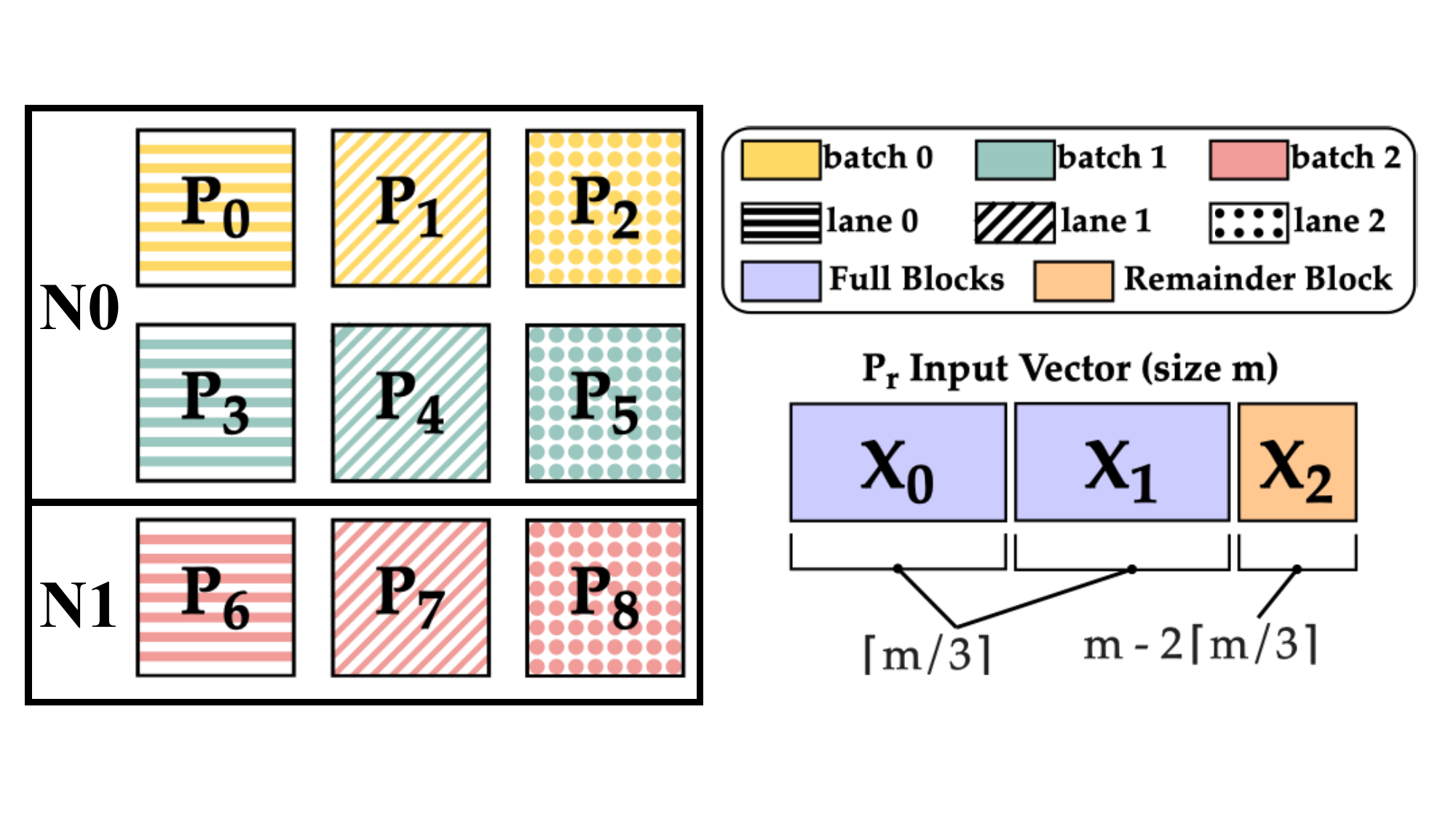}
  \caption{\textbf{Batch--lane topology and staged block mapping.}
Left: $P=9$ (from 2 nodes) and $b=3$ form $B=3$ batches (colors) and $b=3$ lanes (patterns),
with rank $r=xb+i$. Right: each rank partitions $X[0{:}m)$ into $B$ blocks $\{X_d\}$.}
  \label{fig:batch-lane}
  \vspace{-0.5cm}
\end{figure}

We organizes ranks into two logical dimensions, \emph{batch} and \emph{lane}.
Let $P$ be the number of MPI ranks. \tool partitions ranks into batches of  size $b$ (\emph{batch-size}), yielding $B=P/b$ batches. To keep intra-batch communication inside the intended fast domain, $b$ should not larger than $PPN$. Formally, each rank $r\in[0,P)$ is mapped to: $r = xb + i$, where $x$ is the batch ID and $i$ is the lane ID (position within a batch). All ranks with the same 
$i$ across batches form a lane.
This induces a logical $B\times b$
grid: rows are batches and columns are lanes. 
Figure~\ref{fig:batch-lane} (left) illustrates this mapping for $P=9$ and $b=3$. 
Each three ranks in a row with the same color is defined as a \emph{batch}, while each three ranks in a column with the same shape pattern is defined as a \emph{lane}.
Batches are a \emph{logical} partition computed by integer
division and modulo and do not require communicator splits. 

\vspace{-0.3cm}
\subsection{Block and stage decomposition.}
\label{sec:design-new:decomp}

\tool processes the input vector in stages, rather than as a single operation. Each rank contributes a vector $X[0{:}m)$, which is partitioned into $B$ contiguous blocks evenly (one per batch); the last few blocks may contain fewer elements when $m$ is not divisible by $B$ (Figure~\ref{fig:batch-lane} right). At stage $t$, each lane $i$ operates on a  block $d$: $d = t \cdot b + i$.
In each stage, at most one block is assigned per lane, enabling up to $b$ blocks
to be processed with $b$ lanes in parallel. As a result, processing all $B$ blocks
requires $\lceil B/b \rceil$ stages, and the number of stages is
$I = \lceil B/b \rceil = \lceil P/b^2 \rceil$.
This design bounds the active working set that improves data locality, keeps work balanced across lane, reduces global synchronization overhead by decomposing computation into fine-grained steps,  and maximizes parallel utilization of the batch–lane topology.

\begin{figure}[t]
\vspace{-0.3cm}
  \centering
  \includegraphics[width=\columnwidth]{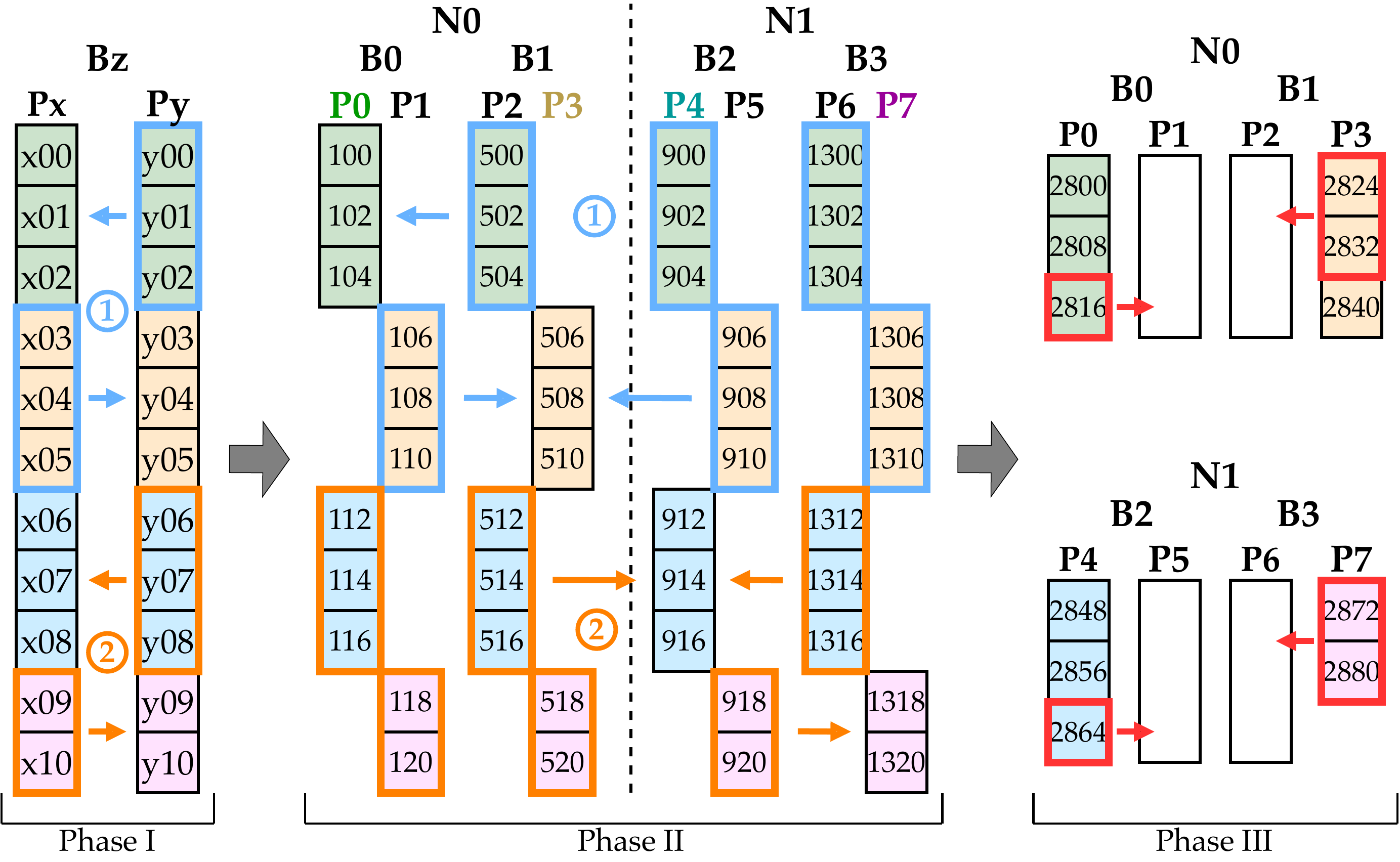}
\caption{Hierarchical Reduce-Scatter follows a three-phase structure. Phase I illustrates intra-batch local reduction using two neighboring ranks, $P_x$ and $P_y$, within a batch $B_z$ containing $m=11$ elements. Phase II captures inter-batch reduction across four batches ($B_0$–$B_3$) spanning two nodes ($N_0$, $N_1$). Phase III completes the process by redistributing the reduced blocks within each batch, ensuring that each block is delivered to its designated owner. Both Phase I and Phase II adopt a staged execution model with two stages (indicated by blue and orange boxes). Color coding consistently tracks blocks according to their destination ranks throughout all phases.}

  \label{fig:rs}
  \vspace{-0.3cm}
\end{figure}
\smallskip
\noindent \textbf{Example:} Consider \(P = 24\) processes with batch size \(b = 3\), yielding \(B = 8\) batches. Suppose each rank \(r\) holds 23 elements. We partition \(X[0\!:\!23)\) into \(8\) blocks, where the first \(7\) blocks contain \(3\) elements each and the last block contains \(2\) elements. In this case, the algorithm proceeds in \(I = \lceil B/b \rceil = 3\) stages, with the stage index \(t\) ranging from \(0\) to \(2\). While the first two stages $(t = 0, 1)$ handle 3 blocks each, the last stage handles 2 blocks. Each batch contains three lanes indexed by \(0,1,2\). At stage \(t\), lane \(i\) operates on block \(d = tb + i\) (whenever \(d < B\)). Thus, stage 0 assigns blocks \(0,1,2\) to lanes \(0,1,2\), respectively; stage 1 assigns blocks \(3,4,5\) to lanes \(0,1,2\); and stage 2 assigns blocks \(6,7\) to lanes \(0,1\).

\subsection{Tunable Parameters}
\label{sec:design-new:tunable}

While the design above defines an efficient communication structure, its performance depends on several key trade-offs, including the balance between intra- and inter- batch communication, and between communication depth and per-round overhead. These trade-offs vary across communication configurations, making a single fixed configuration suboptimal.
To expose and control these trade-offs, \tool introduces a small set of tunable parameters that directly shape the communication structure.

\ke{\textbf{Batch-size:} \tool first introduce a tunable \emph{batch-size} $b$, which controls the number of ranks per batch, determining the granularity of hierarchical decomposition. A larger $b$ increases work performed within fast intra-batch communication, reducing inter-batch communication rounds, but raises intra-batch cost and may exceed fast hardware domain capacity. A smaller $b$ reduces intra-batch overhead at the cost of more global stages and higher synchronization overhead. Thus, $b$ provides a direct mechanism to balance intra-domain efficiency against inter-domain communication cost.}

\ke{\textbf{Intra-batch radices:} \tool further exposes independent radix parameters for intra-batch communication in the Reduce-Scatter and Allgather phases, denoted as $k_{RS}$ and $k_{AG}$. These parameters control the fanout of intra-batch communication. A larger radix reduces communication rounds and synchronization depth but increases per-round volume and contention; a smaller radix has the opposite effect. Crucially, $k_{RS}$ and $k_{AG}$ are tuned independently, reflecting the distinct communication characteristics of each phase.}



\label{sec:design_new}
\begin{figure}[t]
\vspace{-0.3cm}
  \centering
  \includegraphics[width=\columnwidth]{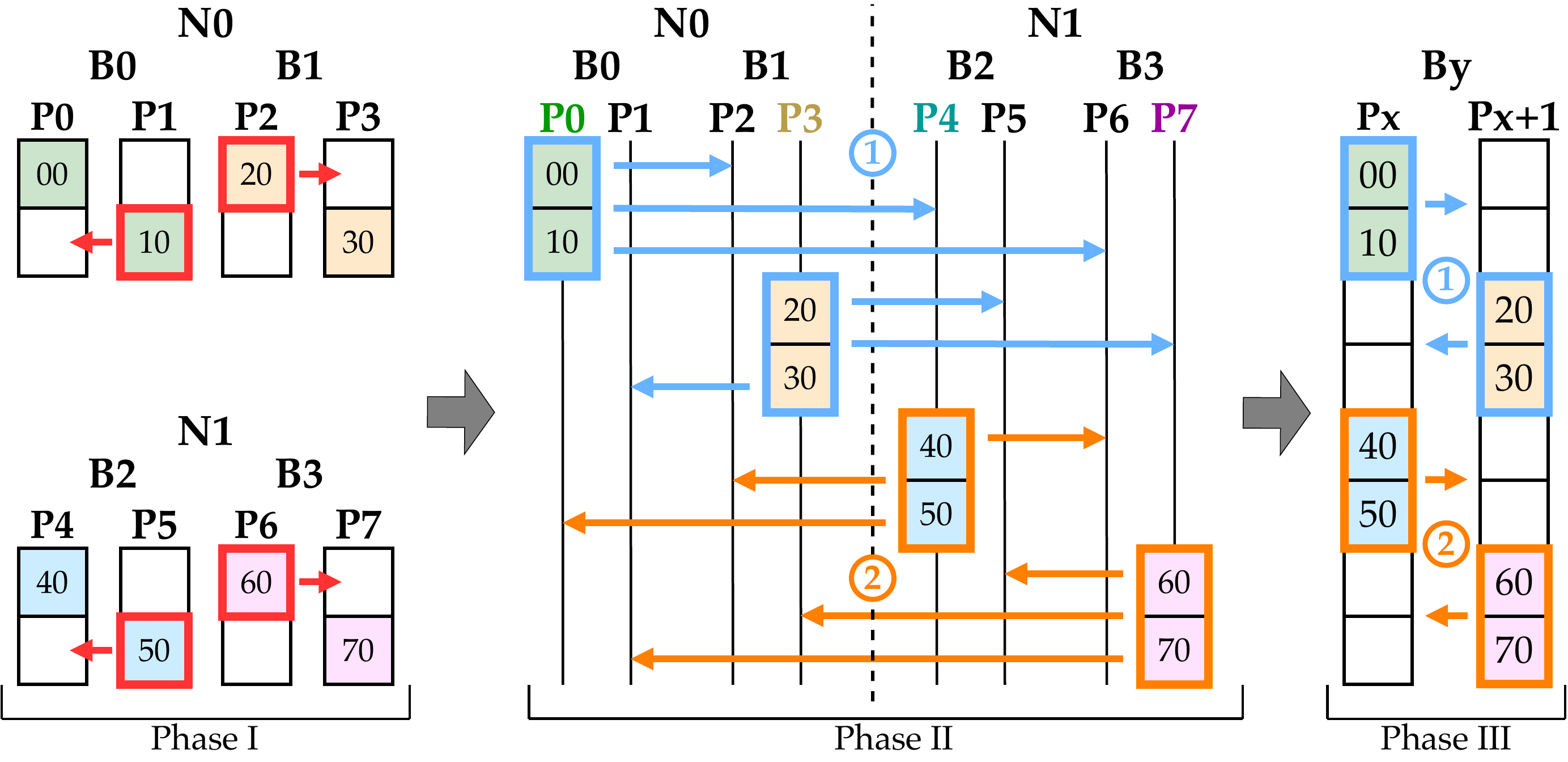}
\caption{\textbf{Hierarchical Allgather (example)} with eight ranks ($P_0$--$P_7$) spanning two nodes ($N_0$, $N_1$) and four batches ($B_0$--$B_3$). Each rank initially holds its scattered partition, with numeric labels denoting example data values. Each stage~$t$ (block $d = tb + i$) proceeds in three phases: \textbf{Phase~I} redistributes blocks within each batch so that each lane rank holds the block assigned to its lane (required only for standalone Allgather); \textbf{Phase~II} broadcasts along lanes across batches via \textsc{LaneOp}(\textsf{BCAST}), with root batch $\rho = tb + i$ sending its block to all other batches; \textbf{Phase~III} replicates all stage blocks within a generic batch~$B_y$ via radix-$k_{\mathrm{AG}}$ intra-batch recursive exchange, so that every rank in the batch holds the complete result. Color coding groups blocks by their originating rank throughout all three phases.}
\label{fig:ag}
\vspace{-0.3cm}
\end{figure}
\section{Proposed Algorithms of \tool}
\label{sec:algo}

\ke{Building on the design in Section~\ref{sec:design-new}, \tool has two hierarchical building blocks: \texttt{Reduce-Scatter} and \texttt{Allgather}. Each of them consists of both intra- and inter-batch communication, following the same \emph{batch-lane} topology and \emph{staged execution} model.
This unified structure enables intra-batch communication to be mapped onto high-bandwidth shared-memory domains, while inter-batch communication is distributed across lanes through a rotating lane-root assignment, allowing multiple ranks to participate concurrently and improving network utilization.}

\ke{In this section, we present the algorithms underlying \tool. We begin with the hierarchical algorithmic structure of the two hierarchical building blocks (Section~\ref{sec:algo:hie}), and then describe their concrete realizations.
For intra-batch communication, \tool designs distinct algorithms for \texttt{Reduce-Scatter} and \texttt{Allgather}, each parameterized by a tunable radix (Section~\ref{sec:algo:intra}). For inter-batch communication, both building blocks share a unified communication primitive, \textsc{LaneOp} (Section~\ref{sec:algo:inter}), which enables fully distributed cross-batch reduction and broadcast.
Finally, \tool introduces a semi-composed execution algorithm that preserves a structured intermediate layout across \texttt{Reduce-Scatter} and \texttt{Allgather}, eliminating redundant data movement between phases (Section~\ref{sec:algo:semi}).}

\subsection{Algorithmic Structure of the Hierarchical Building Blocks}
\label{sec:algo:hie}

\ke{This section outlines the algorithmic structure of the two hierarchical building blocks, \texttt{Reduce-Scatter} and \texttt{Allgather}, followed by their integration via a semi-composed execution that eliminates redundant phases.}

\ke{\textbf{Hierarchical Reduce-Scatter} performs global reduction and produces a distributed partition of the result. It is decomposed into three phases: intra-batch reduction, inter-batch reduction, and intra-batch redistribution. The first phase aggregates partial results within each batch. The second phase completes the reduction across batches to produce globally reduced data. The final phase redistributes the reduced data within each batch to match the output layout required by \texttt{Reduce-Scatter}. Figure~\ref{fig:rs} illustrates this three-phase structure.}

\ke{\textbf{Hierarchical Allgather} replicates distributed data such that every rank obtains the full result. It follows a symmetric three-phase structure: intra-batch alignment, inter-batch broadcast, and intra-batch allgather. The first phase prepares the local data layout for global communication. The second phase disseminates data across batches so that all batches receive the required intermediate results. The final phase completes replication within each batch. Figure~\ref{fig:ag} illustrates this process.}

\begin{figure*}[t]
\vspace{-0.5cm}
  \centering
  \includegraphics[width=0.8\textwidth]{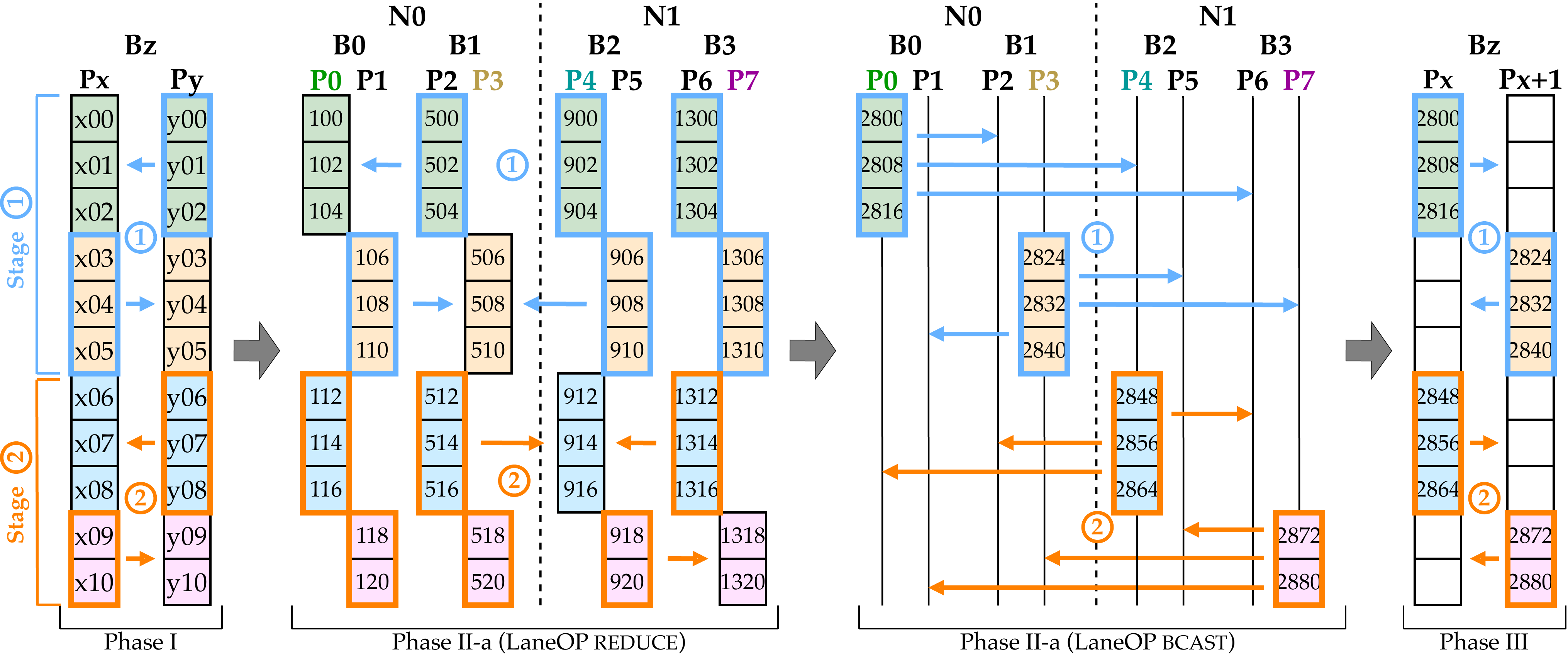}
\caption{\textbf{Semi-composed hierarchical Allreduce example. Allreduce is executed in four phases with this design: intra-batch Reduce-Scatter,   inter-batch lane reduction, inter-batch lane broadcast, and intra-batch Allgather.}}
  \label{fig:allreduce}
\end{figure*}

\ke{\textbf{Semi-composed execution} is then introduced by \tool that preserves a lane-aligned intermediate layout across the boundary of the two building blocks. This design eliminates the need for the intra-batch redistribution phase in \texttt{Reduce-Scatter} and the intra-batch alignment phase in \texttt{Allgather}. As a result, the combined execution proceeds in four phases, reducing redundant data movement and synchronization while remaining consistent with the design described in Section~\ref{sec:design-new:overview}. Figure~\ref{fig:allreduce} illustrates the four-phase.}

\subsection{Intra-batch Algorithms}
\label{sec:algo:intra}

\ke{This section presents the \emph{intra-batch} communication algorithms used in the local phases of both hierarchical building blocks, including intra-batch reduction in \texttt{Reduce-Scatter} and intra-batch allgather in \texttt{Allgather}. Within each batch, communication is performed using \emph{radix}-parameterized recursive exchange, enabling efficient use of shared-memory bandwidth while providing flexible control over communication depth and per-round cost. 
In all algorithms, we assume ranks have been organized into batch-lane topology with $B$ batches and $b$ lanes. The input vector per rank is also partitioned into $B$ contiguous blocks, and and execution proceeds stage by stage (as defined in Section~\ref{sec:design-new}). }

\ke{\textbf{Intra-batch Reduce-scatter:}
Algorithm~\ref{alg:intrars} describes the intra-batch reduction procedure. 
For stage $t$, at most one block $d = tb + i$ is assigned to each lane $i$ (line 1).  
Within each batch, the ranks collaboratively reduce the active blocks using a radix-$k_{\mathrm{RS}}$ recursive-exchange schedule. The radix $k_{\mathrm{RS}}$ determines that each active rank communicates with up to $(k_{\mathrm{RS}} - 1)$ other ranks per round, for a total of $\lceil \log_{k_{\mathrm{RS}}} b \rceil$ rounds.
In each round, ranks exchange only a slice of the active blocks in the current stage for a specific target rank and immediately apply the reduction operator in-place (lines 9-12). 
Radix-$k_{\mathrm{RS}}$ recursive exchange requires the number of participating ranks to be a power of $k_{\mathrm{RS}}$ to form complete communication groups per round.
Thus, when $b$ is not a power of $k_{\mathrm{RS}}$, not all ranks in the batch can participate in the data exchange. In that case, a \emph{non-participating} rank sends its active blocks to a participating rank in the same batch, which is referred to as \emph{fold} (lines 5-6). After the exchange completes, the reduced result corresponding to each \emph{non-participating} rank is sent back, which is referred to as \emph{unfold} (lines 14).
This phase produces a lane-aligned state: for each active lane $i$ in stage $t$, rank $i$ in each batch holds the batch-local reduced result for block $d=tb+i$, thereby combining reduction and local scattering.
}







\begin{algorithm}[t]
\caption{Intra-batch Radix-$k_{\mathrm{RS}}$ Reduce-Scatter}
\footnotesize
\label{alg:intrars}
\begin{algorithmic}[1]
\State $d \gets t b + i$
\If{$d \ge B$}
    \State \Return
\EndIf

\If{non-participating}
    \State send active blocks to a participating rank \Comment{fold}
\EndIf

\State reduce folded data (if any)

\For {each $\lceil \log_{k_{\mathrm{RS}}} b \rceil$ rounds}
    \State exchange slices of active blocks with up to $(k_{\mathrm{RS}} - 1)$ target ranks
    \State apply reduction in-place
\EndFor

\State $U \gets$ block $d$
\State send $U$ to folded ranks (if any) \Comment{unfold}
\State \Return $U$
\end{algorithmic}
\end{algorithm}

\ke{\textbf{Intra-batch Allgather:} 
Algorithm~\ref{alg:intraag} describes the intra-batch allgather procedure.
Due to the same batch-lane topology, this algorithm also enables each lane $i$ handles one reduced block $d$ per stage $t$ (line 1).
Within each batch, the ranks collaboratively replicate the active blocks using a radix-$k_{\mathrm{AG}}$ recursive-exchange schedule, which proceeds in $\lceil \log_{k_{\mathrm{AG}}} b \rceil$ rounds (line 5). In each round, a rank communicates with up to $(k_{\mathrm{AG}} - 1)$ peers (line 6). Specifically, for each neighbor, the rank receives a contiguous segment of blocks and sends out its currently available prefix of blocks (lines 5–9). After each round, the number of blocks available at a rank increases by a factor of $k_{\mathrm{AG}}$ (line 8).
When $b$ is not a power of $k_{\mathrm{AG}}$, some communication groups in a round may be incomplete. In that case, the exchanged segment size is clipped to avoid exceeding the valid block range (line 7).
After all rounds complete, each rank holds all active blocks for the current stage. A final local reindexing step restores the blocks to their natural lane order (line 10), completing the intra-batch allgather.}



\subsection{Inter-batch Algorithm (LaneOp)}
\label{sec:algo:inter}

Both hierarchical Reduce-Scatter and hierarchical Allgather require inter-batch communication. \tool expresses all such inter-batch traffic through
\textsc{LaneOp} (Algorithm~\ref{alg:laneop}), which is invoked in two
complementary modes: \textsf{REDUCE} for lane-wise accumulation (used by
Reduce-Scatter) and \textsf{BCAST} for lane-wise distribution (used by Allgather).

Formally, for each stage $t$ and lane $i$, \textsc{LaneOp} selects a unique
\emph{root batch} $\rho = tb + i$ (line~\ref{ln:laneop-rho}).
If $\rho \ge B$, lane $i$ is inactive and \textsc{LaneOp} returns immediately.
Otherwise, the corresponding \emph{lane root rank} is $root = \rho b + i$ (line~\ref{ln:laneop-root}).
In \textsf{REDUCE} mode, the root rank accumulates values across batches: it initializes the accumulator with its local value and receives contributions from all other batches along the same lane, applying the reduction operator upon arrival (lines~\ref{ln:laneop-mode-reduce}--\ref{ln:laneop-reduce-recv}).
Non-root ranks simply send their local value to $root$ (line~\ref{ln:laneop-reduce-send}).
In \textsf{BCAST} mode, the communication pattern is reversed: the root rank sends its value to all other batches, while non-root ranks receive from $root$ (lines~\ref{ln:laneop-mode-bcast}--\ref{ln:laneop-bcast-recv}).
Because $\rho$ depends on $(t,i)$, the root responsibility rotates across lanes and stages, distributing inter-batch communication across ranks rather than concentrating it on a fixed leader.

\begin{algorithm}[t]
\caption{Rotating-root inter-batch lane operation}
\label{alg:laneop}
\footnotesize
\begin{algorithmic}[1]
\Require mode $\in\{\textsf{REDUCE},\textsf{BCAST}\}$, stage $t$, batch id $x$, lane id $i$
\Require $B$ batches, batch size $b$, local block $V$ (size $m/B$), reduction $\oplus$
\Ensure updated block $V'$ (reduced at root / broadcast to all)
\State $\rho \gets tb + i$ \Comment{root batch} \label{ln:laneop-rho}
\If{$\rho \ge B$} \State \Return $V$ \EndIf
\State $root \gets \rho\cdot b + i$ \Comment{lane root rank} \label{ln:laneop-root}

\If{mode $=$ \textsf{REDUCE}} \label{ln:laneop-mode-reduce}
  \If{$x=\rho$} \Comment{root accumulates} \label{ln:laneop-reduce-root}
    \State $A \gets V$
    \For{$y=0$ to $B-1$, $y\neq x$}
      \State recv $T$ from rank $(y\cdot b + i)$; \ $A \gets A \oplus T$ \label{ln:laneop-reduce-recv}
    \EndFor
    \State \Return $A$
  \Else
    \State send $V$ to rank $root$; \State \Return $V$ \label{ln:laneop-reduce-send}
  \EndIf
\Else \Comment{mode $=$ \textsf{BCAST}} \label{ln:laneop-mode-bcast}
  \If{$x=\rho$} \Comment{root distributes} \label{ln:laneop-bcast-root}
    \For{$y=0$ to $B-1$, $y\neq x$}
      \State send $V$ to rank $(y\cdot b + i)$ \label{ln:laneop-bcast-send}
    \EndFor
  \Else
    \State recv $V$ from rank $root$ \label{ln:laneop-bcast-recv}
  \EndIf
  \State \Return $V$
\EndIf
\end{algorithmic}
\end{algorithm}

\begin{algorithm}[t]
\caption{Intra-batch Radix-$k_{\mathrm{AG}}$ Allgather}
\footnotesize
\label{alg:intraag}
\begin{algorithmic}[1]
\State $d \leftarrow tb + i$
\If{$d \ge B$} \Return \EndIf

\State $G \leftarrow \{\text{block } d\}$

\For{each $\lceil \log_{k_{\mathrm{AG}}} b \rceil$ rounds}
    \State exchange subsets of $G$ with up to $(k_{\mathrm{AG}} - 1)$ neighbors
    \State clip exchanged blocks for non-powers of $k_{\mathrm{AG}}$
    \State expand $G$ by a factor of $k_{\mathrm{AG}}$
\EndFor

\State $U \gets$ block $d$
\State reorder $G$ to natural block order \Comment{unfold}
\State \Return $G$
\end{algorithmic}
\end{algorithm}

\subsection{Semi-composed Execution Algorithm}
\label{sec:algo:semi}

\ke{The above building blocks can be composed to implement \texttt{Allreduce}. However, a naive composition introduces redundant intra-batch work at the boundary between \texttt{Reduce-Scatter} and \texttt{Allgather}.
In particular, hierarchical Reduce-Scatter performs a batch-local redistribution to place reduced blocks on their final scattered owners, while hierarchical Allgather begins with a batch-local lane-alignment step to reconstruct the layout required for lane-wise broadcast.}

\ke{\tool avoids both these redundant steps by preserving the lane-aligned intermediate layout, as shown in Algorithm~\ref{alg:chiara_allreduce}. 
In Phase I (lines 4–9), each batch performs an intra-batch Reduce-Scatter. Each batch locally reduces the corresponding blocks using radix-$k_{\mathrm{RS}}$ recursive exchange, producing a lane-aligned partial result $R_t$.
In Phase II-a (lines 10–12), the reduction is completed across batches. For each stage, \textsc{LaneOp} is invoked in \textsf{REDUCE} mode to aggregate the lane-aligned values across batches, producing the globally reduced block $U_t$ at the lane root.
In Phase II-b (lines 13–15), the reduced result is distributed back to all batches. \textsc{LaneOp} is invoked in \textsf{BCAST} mode so that each batch obtains the reduced block at its corresponding lane rank $(x,i)$, preserving the lane-aligned layout.
In Phase III (lines 16–18), each batch performs an intra-batch Allgather using radix-$k_{\mathrm{AG}}$ recursive exchange. Starting from the lane-aligned blocks, each rank replicates the stage blocks within the batch, producing the final output segments $Y_t$.}

\ke{After Phase II-b, each batch already holds the reduced stage blocks in the exact layout required by Phase III, enabling a direct composition of Reduce-Scatter Phases I–II with Allgather Phases II–III without any intermediate local reorganization. Importantly, this semi-composition does not constrain the intra-batch algorithms: the reduction and replication phases remain independent, exposing separate radices $k_{\mathrm{RS}}$ and $k_{\mathrm{AG}}$, while sharing the same inter-batch lane schedule.}

\begin{algorithm}[t]
\caption{\tool: Semi-composed hierarchical Allreduce}
\label{alg:chiara_allreduce}
\footnotesize
\begin{algorithmic}[1]
\Require rank $r$, size $P$, input $X[0:m)$, op $\oplus$, batch $b$, radixes $k_{AG}, k_{RS}$
\Ensure output $Y[0:m) = \bigoplus_{p=0}^{P-1} X_p$
\State $i \gets r \bmod b$; \ $x \gets \lfloor r/b \rfloor$ \Comment{lane ID, batch id}
\State $B \gets P/b$; \ $I \gets \lceil B/b \rceil$ \Comment{$=\lceil P/b^2\rceil$}
\State split $X$ into $B$ equal blocks $\{X_d\}_{d=0}^{B-1}$ (each size $m/B$)

\For{$t=0$ to $I-1$} \Comment{Phase I: intra-batch Reduce-Scatter}
  \State $d \gets t\cdot b + i$
  \If{$d < B$}
    \State $R_t \gets \textsc{IntraRS}_k\!\left(i,b,k_{RS},\{X_{t b},\dots,X_{\min((t+1)b,B)-1}\},\oplus\right)$
  \EndIf
\EndFor

\For{$t=0$ to $I-1$} \Comment{Phase II-a: inter-batch lane reduction}
  \State $U_t \gets \textsc{LaneOp}(\textsf{REDUCE},t,x,i,B,b,R_t,\oplus)$
\EndFor

\For{$t=0$ to $I-1$} \Comment{Phase II-b: inter-batch lane broadcast}
  \State $R_t \gets \textsc{LaneOp}(\textsf{BCAST},t,x,i,B,b,U_t,\oplus)$
\EndFor

\For{$t=0$ to $I-1$} \Comment{Phase III: intra-batch Allgather (radix $k$)}
  \State $Y_t \gets \textsc{IntraAG}_k(i,b,k_{AG},R_t)$ \Comment{$b$ blocks for stage $t$}
\EndFor
\end{algorithmic}
\end{algorithm}


\section{Evaluation}
\label{sec:eval}

We evaluate \tool using microbenchmarks on three leadership-class systems:
Polaris and Aurora at \ac{ALCF}, and Fugaku at RIKEN R-CCS. Polaris is an HPE Cray
Slingshot~11 system with single-socket AMD EPYC (Zen3) CPUs (32 cores per node),
Aurora is an HPE Cray Slingshot~11 system with dual-socket Intel Xeon Max CPUs
(104 cores per node), and Fugaku uses Fujitsu A64FX CPUs (48 user-accessible cores
per node) interconnected by the Tofu-D 6D mesh/torus network. All measurements
use CPU-resident MPI collectives and run at full node concurrency (one rank per
core). Since \tool uses a \emph{logical} batch mapping based on contiguous rank
ids (Section~\ref{sec:design-new}), ranks are placed so that contiguous rank groups
map to the intended locality domain (socket on multi-socket nodes, node on
single-socket systems). Each reported runtime is the median over 50 iterations.
The evaluation follows the design structure in Section~\ref{sec:design-new}. We first
study the batch-size parameter \(b\) (Section~\ref{sec:eval-b}), which defines the
batch--lane decomposition and controls the number of inter-batch stages
\(I=\lceil P/b^2\rceil\). We then examine radix selection for the hierarchical
building blocks (Section~\ref{sec:eval-building-blocks}), focusing on the distinct
behaviors of \(k_{\mathrm{AG}}\) for Allgather and \(k_{\mathrm{RS}}\) for
Reduce-Scatter. Next, we quantify the end-to-end benefit of semi-composition in
\tool (Section~\ref{sec:eval-allreduce}). Finally, we compare \tool against
vendor \texttt{MPI\_Allreduce} implementations on all three systems
(Section~\ref{sec:eval-vendor}).

\smallskip
\noindent\textbf{Message-size convention.}
Unless stated otherwise, all plots report the \emph{per-rank segment size} $S$
(in doubles) used by the staged schedule. This $S$ corresponds to the Allgather
send count (equivalently, the Reduce-Scatter receive count). The equivalent
\texttt{MPI\_Allreduce} element count is $P\cdot S$.

\begin{figure}[t]
  \vspace{-0.5cm}
  \centering
  \includegraphics[width=\columnwidth,trim=0cm .8cm 0cm 0cm,clip]{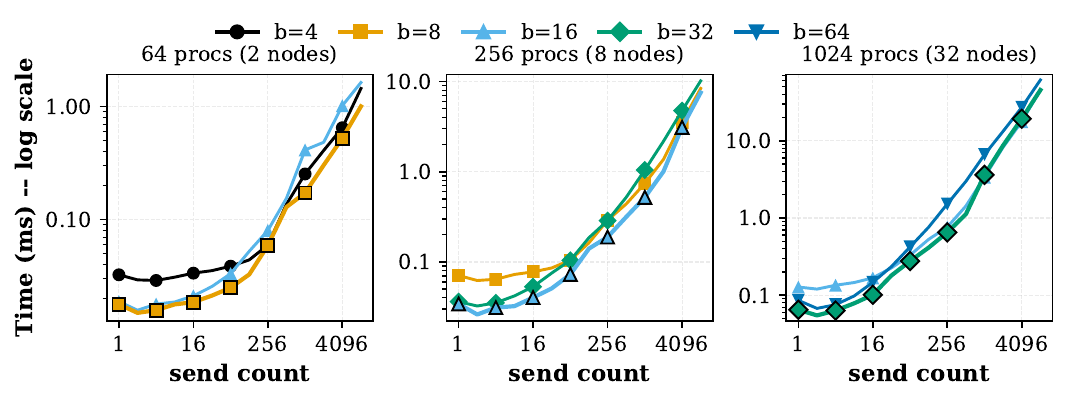}
  \vspace{-7pt}%
  \hrule height 0.4pt
  \vspace{0pt}%
  \includegraphics[width=\columnwidth]{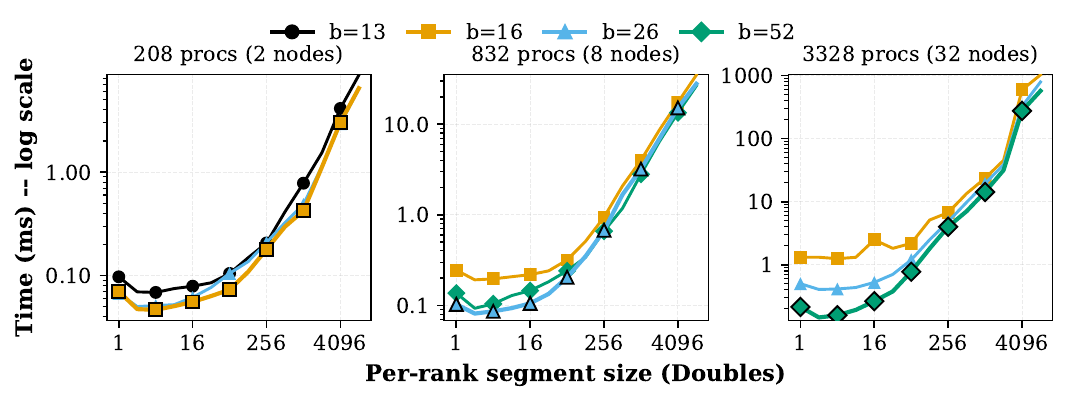}
  \caption{Sensitivity of hierarchical Allgather runtime to the batch size \(b\)
  on Polaris (top) and Aurora (bottom).}
  \label{fig:eval-batch}
  \vspace{-0.5cm}
\end{figure}
\subsection{Batch Decomposition}
\label{sec:eval-b}

The batch size \(b\) defines the logical batch--lane topology in
Section~\ref{sec:design-new}: ranks are grouped into contiguous batches of size \(b\),
and inter-batch communication occurs along lanes in \(I=\lceil P/b^2\rceil\)
stages. Increasing \(b\) reduces the number of stages, with \(I=1\) once
\(b \ge \sqrt{P}\). For \(b < \sqrt{P}\), multiple full stages are required and
the final stage may be partial. At the same time, larger \(b\) increases
the cost of the intra-batch phases because more ranks participate in each
recursive-exchange round. To keep intra-batch communication inside the intended
fast domain, we additionally restrict \(b \le b_{\max}\), where \(b_{\max}\)
is the maximum number of ranks placed on a single socket (and equals ranks per
node on single-socket systems).

Figure~\ref{fig:eval-batch} shows the sensitivity of hierarchical Allgather
runtime to \(b\) on Polaris (top) and Aurora (bottom) for representative process
counts. Two trends are consistent across configurations. First, the best-performing
\(b\) lies near \(\sqrt{P}\), reflecting the reduction in inter-batch stages
\(I=\lceil P/b^2\rceil\). Second, choosing \(b\) beyond the socket limit degrades
performance because the intra-batch phases spill into slower locality levels; this
effect is visible on systems where candidate \(b\) values exceed \(b_{\max}\).
Across segment sizes, the best-performing \(b\) is stable and its neighbors
are the next-best choices, with runtimes degrading as \(b\)
moves away from the optimum.

Based on these results, all subsequent experiments select \(b\) using a simple
rule consistent with the cost-model guidance: among divisors of \(P\), choose the
value \(b \le b_{\max}\) that is closest to \(\sqrt{P}\) (breaking ties toward the
larger \(b\)). This behavior holds for Reduce-Scatter since it
shares the same stage structure and inter-batch schedule.


\begin{figure}[t]
\vspace{-0.5cm}
  \centering
  {\small Polaris}

  \includegraphics[width=\columnwidth,trim=0.27cm .7cm 0.25cm 0cm,clip]{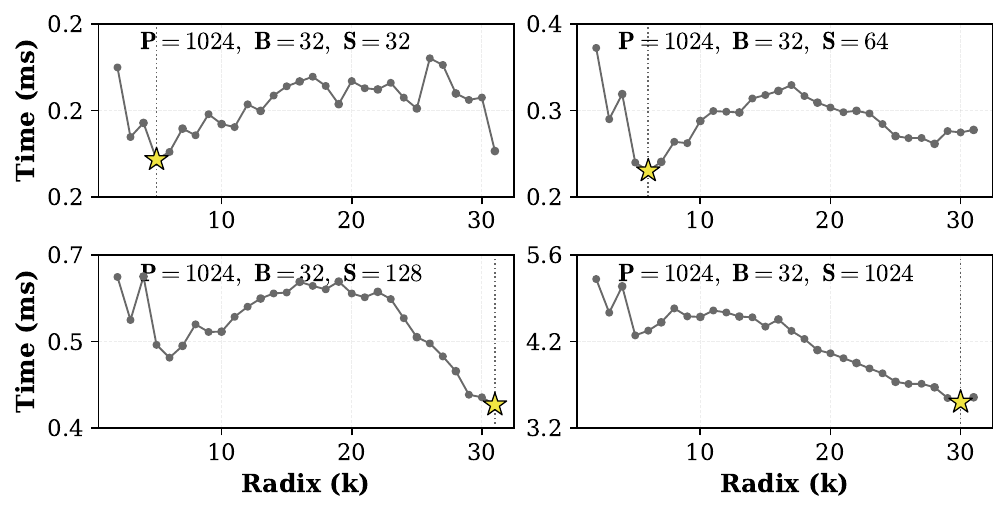}
  \vspace{-10pt}%
  \hrule height 0.4pt
  \vspace{5pt}%
  {\small Aurora}

  \includegraphics[width=\columnwidth,trim=0.27cm .7cm 0.25cm 0cm,clip]{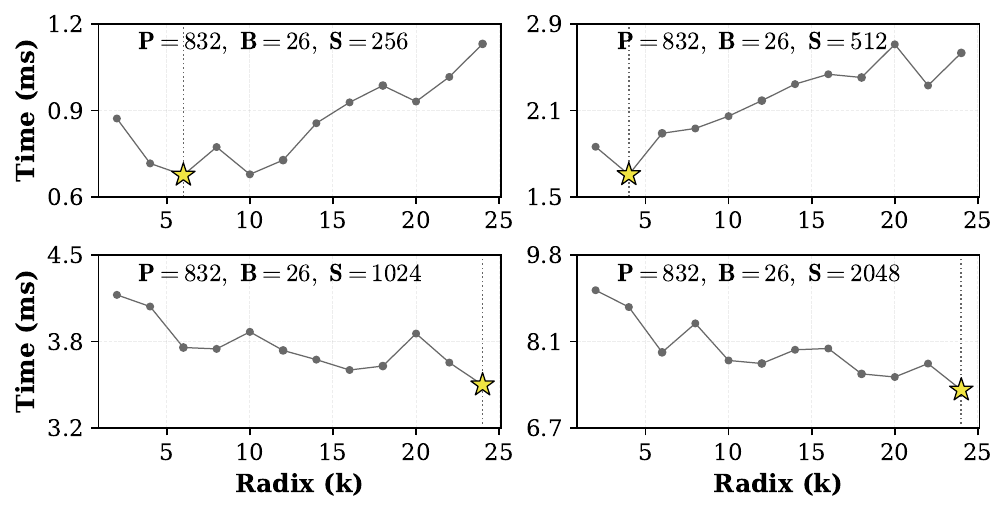}
  \vspace{-10pt}%
  \hrule height 0.4pt
  \vspace{5pt}%
  {\small Fugaku}

  \includegraphics[width=\columnwidth,trim=0.27cm 0cm 0.25cm 0cm,clip]{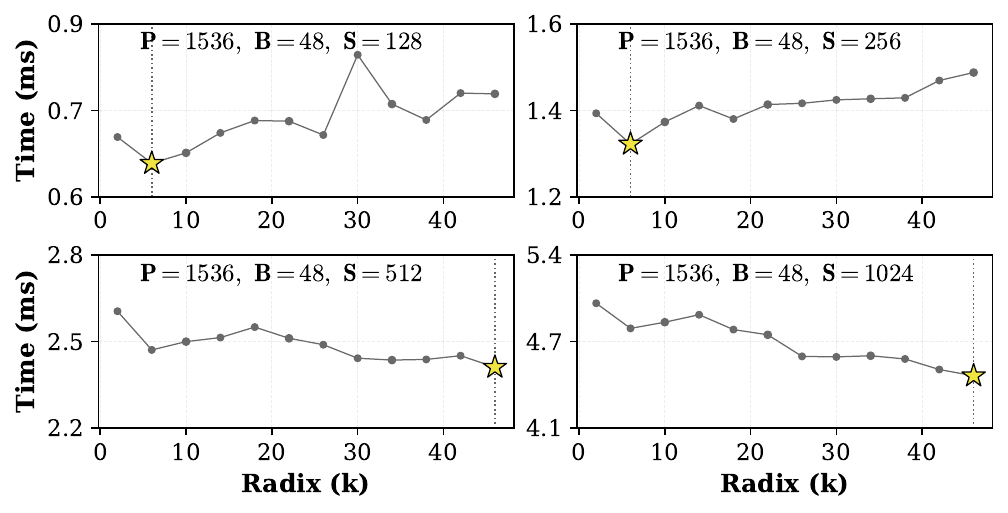}

  \caption{Radix sensitivity of hierarchical Allgather on different systems. Execution time is shown as a function of \(k_{\mathrm{AG}}\) for
  representative process counts and and per-rank segment sizes.}
  \label{fig:eval-ag-radix}
  \vspace{-0.5cm}
\end{figure}

\begin{figure}[t]
\vspace{-0.5cm}
  \centering
  Polaris

  \includegraphics[width=\columnwidth,trim=0.27cm .7cm 0.25cm 0cm,clip]{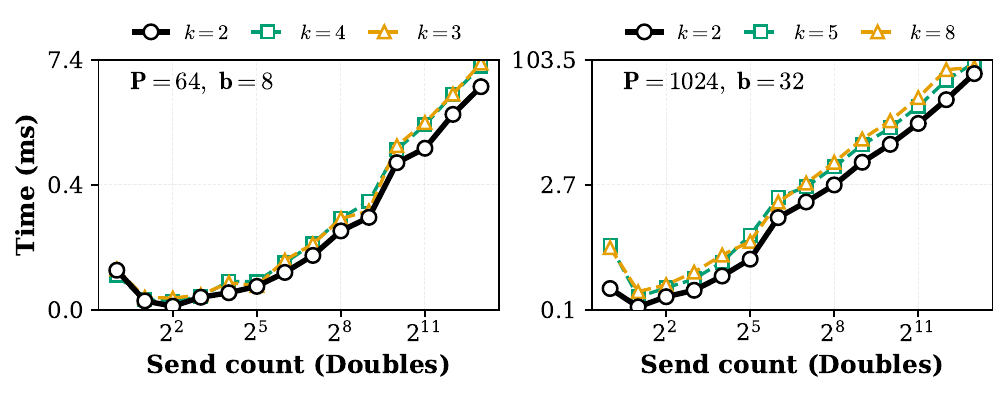}
  \vspace{-7pt}%
  \hrule height 0.4pt
  \vspace{5pt}%
  Aurora

  \includegraphics[width=\columnwidth,trim=0.27cm .7cm 0.25cm 0cm,clip]{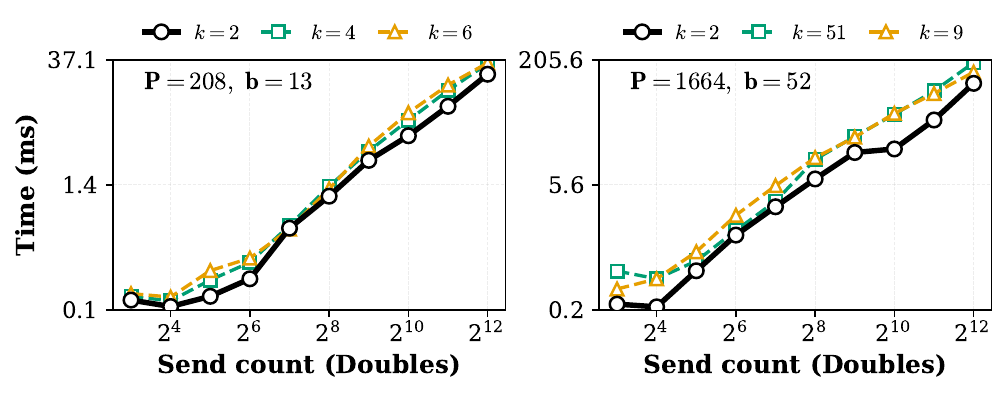}
  \vspace{-7pt}%
  \hrule height 0.4pt
  \vspace{5pt}%
  Fugaku

  \includegraphics[width=\columnwidth,trim=0.27cm 0cm 0.21cm 0cm,clip]{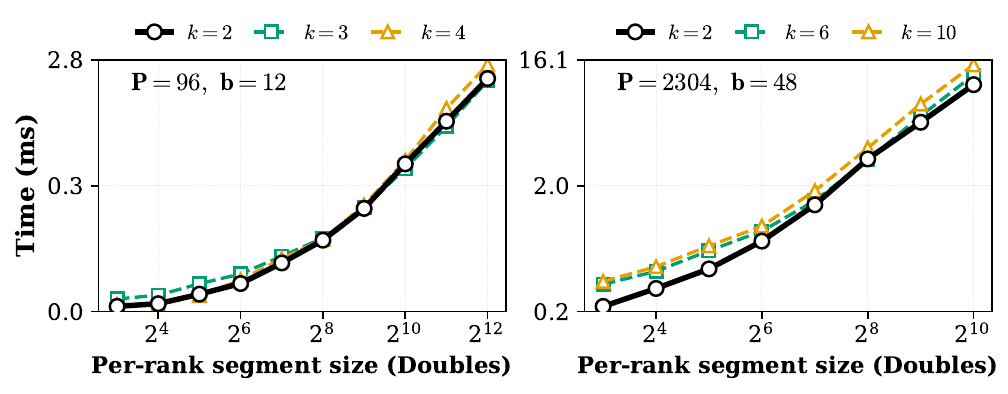}
  \caption{Radix sensitivity of hierarchical Reduce-Scatter. Each panel shows the
  three best-performing \(k_{\mathrm{RS}}\) values (selected by overall runtime
  across per-rank segment sizes) for a small and a large configuration on each system. In all
  cases, the binary radix \(k_{\mathrm{RS}}=2\) is optimal across segment sizes.}
  \label{fig:eval-rs-radix}
  \vspace{-0.3cm}
\end{figure}

\subsection{Radix Sensitivity of the Building Blocks}
\label{sec:eval-building-blocks}

We now examine how the intra-batch radices affect the two hierarchical building
blocks. Recall that \tool exposes \emph{two} independent radices:
\(k_{\mathrm{RS}}\) for the intra-batch Reduce-Scatter phase and
\(k_{\mathrm{AG}}\) for the intra-batch Allgather phase. This separation matters
because the two collectives stress different parts of the communication stack and
have different dependency structures.

Figure~\ref{fig:eval-ag-radix} sweeps \(k_{\mathrm{AG}}\) for \textit{hierarchical Allgather}
on all three systems. For small segment sizes, the best-performing radix consistently
clusters near \(k_{\mathrm{AG}} \approx \sqrt{b}\). In this latency-dominated regime,
a moderate fanout balances the reduction in communication depth against per-round
synchronization and message-matching overhead. As the per-rank segment size increases, the
optimum shifts toward larger radices, and for medium to large messages the minimum
is attained near the upper end of the admissible range, with \(k_{\mathrm{AG}}\)
often close to \(b-1\). This trend reflects the recursive-exchange trade-off: larger \(k_{\mathrm{AG}}\)
reduces the number of rounds \(\lceil \log_{k_{\mathrm{AG}}} b\rceil\) but
increases per-round fanout and data per round. For larger messages, the optimum
therefore depends on whether the platform can sustain the higher fanout; in our
measurements it typically shifts toward \(k_{\mathrm{AG}}\) close to \(b-1\).

Reduce-Scatter shows markedly different behavior. Figure~\ref{fig:eval-rs-radix}
plots the three best-performing radices for hierarchical Reduce-Scatter on each system.
On all platforms and per-rank segment sizes, the radix \(k_{\mathrm{RS}}=2\)
consistently minimizes runtime. Intuitively, the reduction dependency structure
limits the benefit of aggressive fan-in, while larger radices increase per-round
message pressure and local reduction overhead without sufficient depth reduction to
compensate. Consequently, we fix \(k_{\mathrm{RS}}=2\) in the remainder of the
evaluation and treat \(k_{\mathrm{AG}}\) as the primary message-size-dependent
tuning knob.

\begin{figure}[t]
  \centering
  \includegraphics[width=0.8\columnwidth]{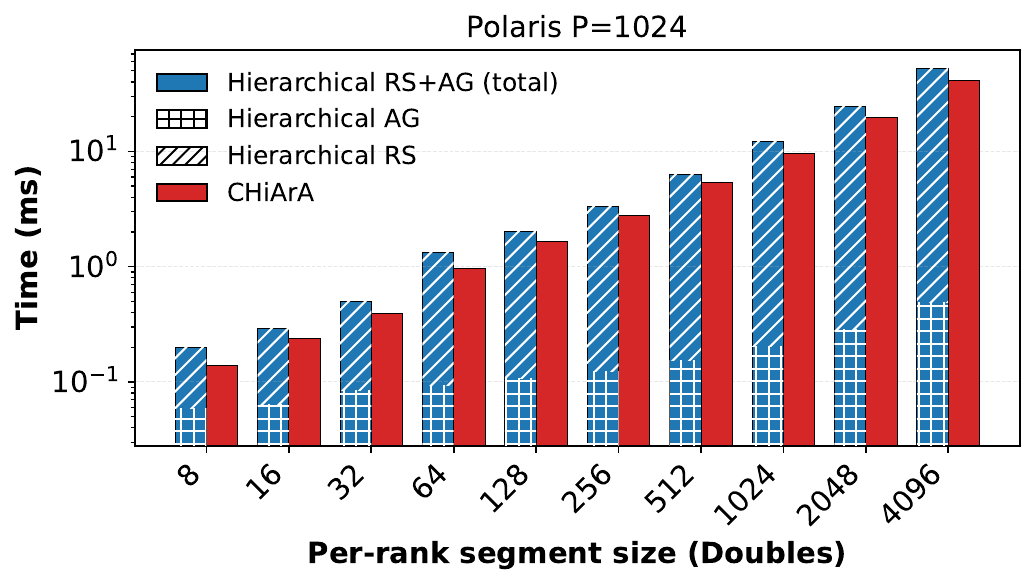}
  \caption{Effect of semi-composition on Allreduce runtime.}
  \label{fig:eval-join}
  \vspace{-0.5cm}
\end{figure}

\subsection{Impact of Semi-Composition}
\label{sec:eval-allreduce}

We next quantify the benefit of semi-composition in \tool. As shown in
Section~\ref{sec:algo:semi}, \tool preserves the lane-aligned intermediate
layout produced by inter-batch reduction and reuses it as the input to the
inter-batch broadcast and the final intra-batch replication. This removes the
redundant boundary work that would otherwise occur when composing standalone
Reduce-Scatter and standalone Allgather.

Figure~\ref{fig:eval-join} compares \tool against a naive hierarchical Allreduce
constructed as standalone Reduce-Scatter followed by standalone Allgather, and also
shows the corresponding phase components. Across segment sizes, semi-composition
consistently reduces runtime by eliminating redundant intra-batch data movement
and synchronization at the Reduce-Scatter to Allgather boundary, with the largest
relative gains appearing in the small to medium regime where intra-batch overheads
are most visible.

\begin{figure}[t]
\vspace{-0.5cm}
  \centering
\includegraphics[width=\columnwidth]{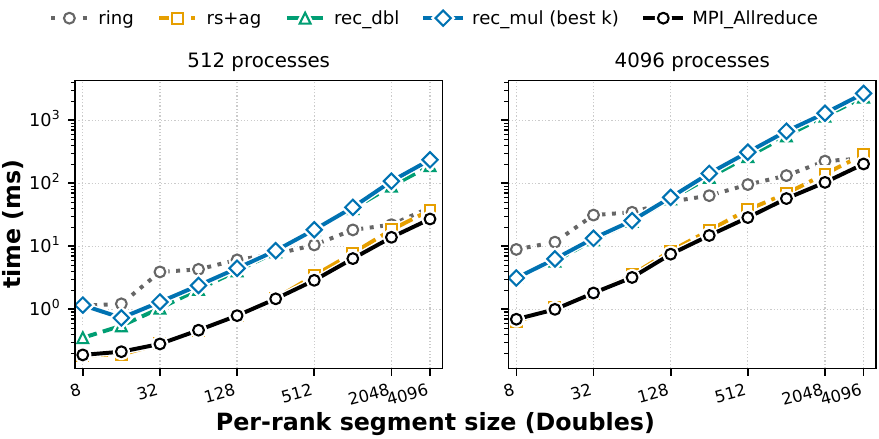}
\caption{Allreduce latency versus per-rank segment size for 512 and 4096 processes. The default \texttt{MPI\_Allreduce} in MPICH closely tracks the fastest internal algorithm across message sizes, motivating its use as the baseline.}
  \label{fig:eval-mpich}
  \vspace{-0.5cm}
\end{figure}

\begin{figure*}[t]
  \centering
  \vspace{-0.5cm}
  \includegraphics[width=0.33\linewidth]{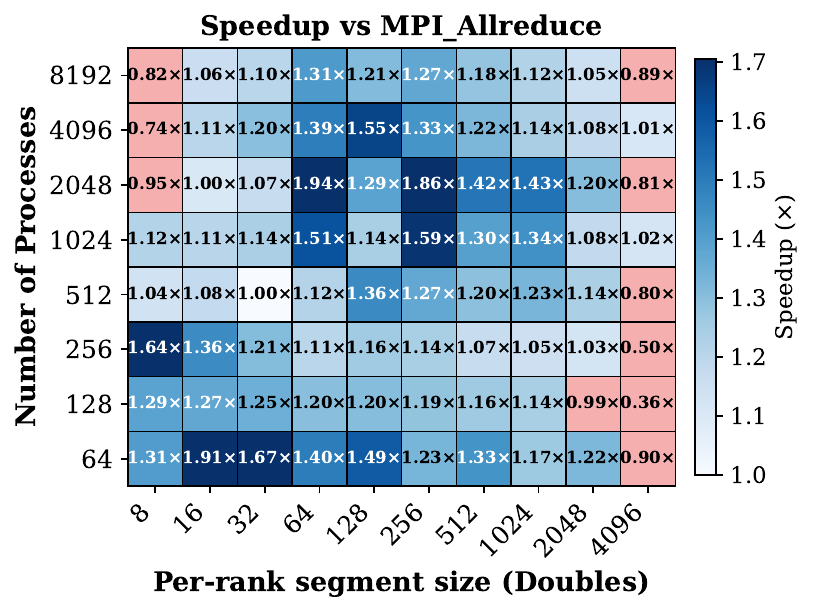}
  \includegraphics[width=0.31\linewidth]{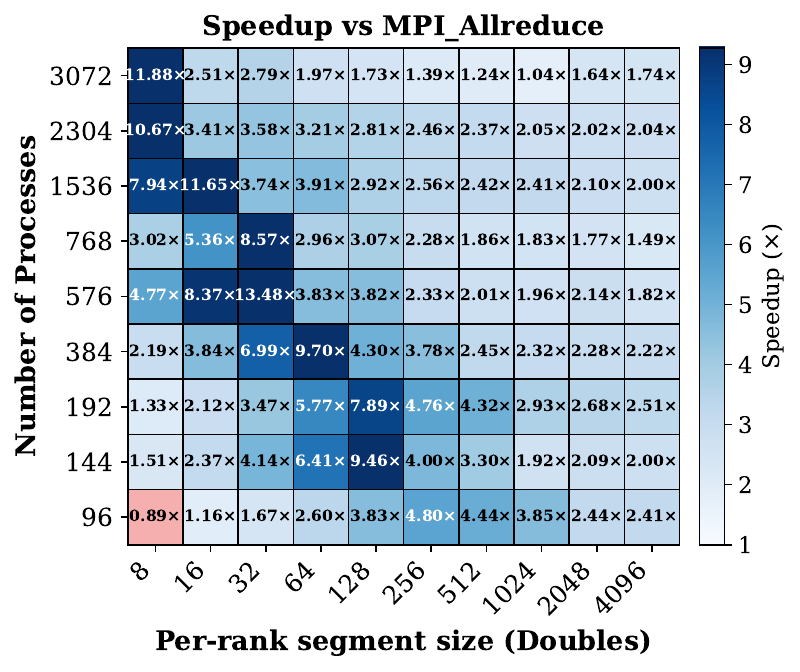}
  \includegraphics[width=0.33\linewidth]{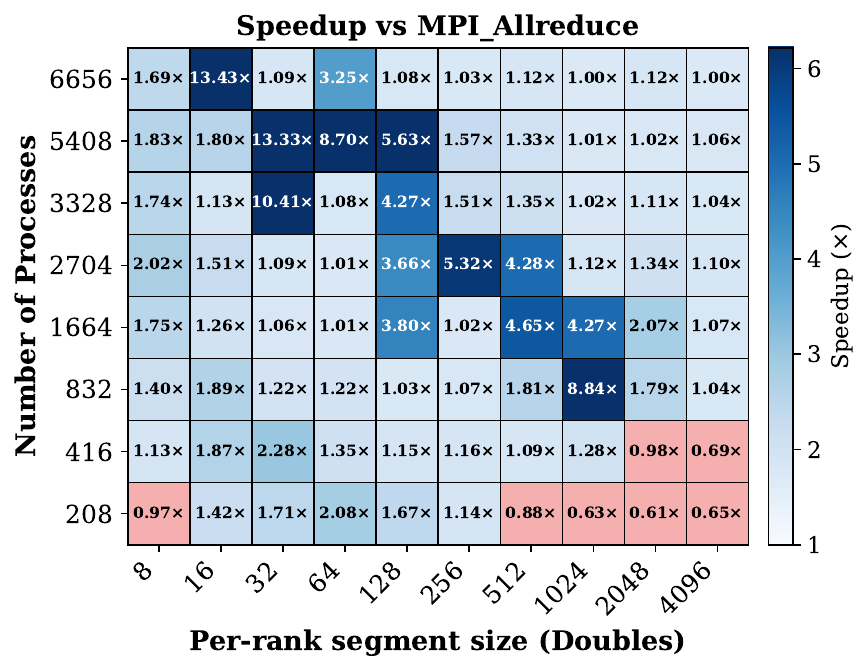}
  
  \caption{Speedup of \tool relative to vendor \texttt{MPI\_Allreduce} on Polaris (left), fugaku (center) and Aurora (right).}
  \label{fig:eval-perf-pol}
  \vspace{-0.3cm}
\end{figure*}



\subsection{Comparison with Vendor \texttt{MPI\_Allreduce}}
\label{sec:eval-vendor}

We now compare the final \tool configuration against the vendor
\texttt{MPI\_Allreduce} on each platform. All \tool parameters are fixed according
to the sensitivity studies above: \(b\) is selected by the batch rule in
Section~\ref{sec:eval-b}, \(k_{\mathrm{RS}}\) is fixed to 2 based on
Figure~\ref{fig:eval-rs-radix}, and \(k_{\mathrm{AG}}\) is selected based on the
message-size trends observed in Figure~\ref{fig:eval-ag-radix}. We use vendor
\texttt{MPI\_Allreduce} as the baseline because it performs internal algorithm
selection as a function of per-rank segment size and scale; Figure~\ref{fig:eval-mpich}
explicitly validate that this baseline tracks the best available
MPICH variants across the studied regimes.

Figures~\ref{fig:eval-perf-pol} report the speedup of
\tool relative to the baseline as a function of process count and per-rank
segment size. Each heatmap cell reports the ratio between the median runtime of the vendor
\texttt{MPI\_Allreduce} and that of \tool for the same configuration; values larger
than 1 indicate configurations where \tool is faster.

On Polaris (Figure~\ref{fig:eval-perf-pol} (left)), \tool provides consistent gains in
the small to medium regime and at scale, reaching up to about \(1.9\times\).
Aurora (Figure~\ref{fig:eval-perf-pol} (right)) shows substantially larger improvements,
with speedups exceeding \(10\times\) in several regions and peaking above
\(13\times\), reflecting the deeper effective hierarchy and higher sensitivity to
synchronization depth at scale. Fugaku (Figure~\ref{fig:eval-perf-pol} (center)) shows a
similar pattern, with the strongest gains concentrated in small to medium send
counts and moderate to large process counts. For very large messages, speedups
approach 1 as bandwidth dominates and baseline implementations approach peak
throughput.


\section{Application Case Study: Parallel K-Means}
\label{sec:app-kmeans}

\begin{figure}[h]
  \centering
  \includegraphics[width=0.49\columnwidth]{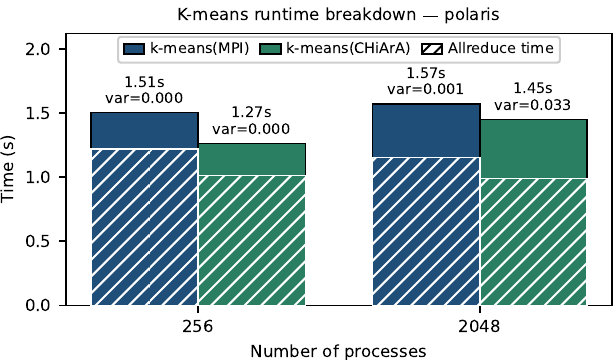}\hfill
  \includegraphics[width=0.49\columnwidth]{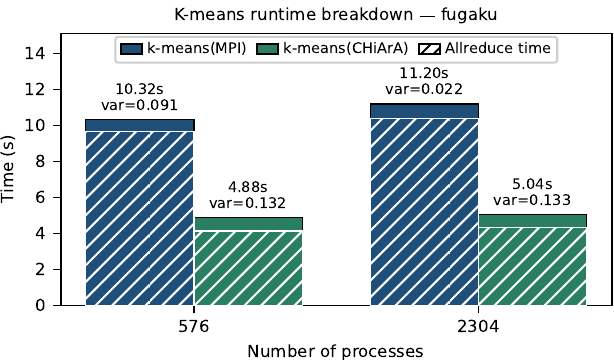}
  \caption{\textbf{End-to-end impact of \tool in Parallel K-Means.}
  Runtime breakdown for two scales on Polaris (left) and Fugaku (right), comparing
  the vendor \texttt{MPI\_Allreduce} against \tool for all Allreduce calls in the
  iteration loop. Stacked bars show the iteration-phase wall-clock time on the
  critical path; the hatched portion is the cumulative time spent inside the three
  Allreduce operations per iteration (counts, centroid sums, and convergence), and
  the remaining portion is local compute and other MPI overheads. Numbers above
  bars report the mean runtime and the observed run-to-run variance.}
  \label{fig:kmeans-minmax-polaris-fugaku}
  \vspace{-0.5cm}
\end{figure}

To complement the microbenchmarks in Section~\ref{sec:eval}, we evaluate \tool in
an end-to-end application by integrating it into an open-source parallel
MPI+OpenMP implementation of Lloyd-style $k$-means clustering\cite{scarlatti}. In each iteration,
every rank (i) assigns its local points to the closest centroid, (ii) accumulates
per-cluster membership counts and per-cluster coordinate sums, and (iii) forms
updated global centroids via collective reductions. Concretely, the implementation
invokes \texttt{MPI\_Allreduce} three times per iteration: a reduction of the
cluster counts (length $K$ integers), a reduction of the centroid sums (length
$K\cdot D$ doubles), and a scalar reduction used for convergence detection. The
centroid-sum Allreduce is the dominant communication in this workload, and it is
exactly the pattern targeted by \tool: repeated dense Allreduces on CPU-resident
buffers inside a bulk-synchronous loop.

We integrated \tool as a drop-in replacement by introducing a wrapper that preserves \texttt{MPI\_Allreduce} semantics for
the datatypes and operator used by the application (here, \texttt{MPI\_INT} and
\texttt{MPI\_DOUBLE} with \texttt{MPI\_SUM}) and dispatches at runtime either to
the vendor \texttt{MPI\_Allreduce} or to \tool. This isolates the effect of the
collective implementation: the computation, data distribution, and convergence
logic are identical across runs. \tool is configured using the same
parameter-selection strategy derived in Section~\ref{sec:eval}.

We run synthetic instances to ensure controlled, repeatable communication volume
while exercising the full iterative path. Each iteration performs three Allreduces:
(i) a cluster-count reduction of $K$ integers, (ii) a centroid-sum reduction of
$K\cdot D$ doubles, and (iii) a convergence reduction of one integer. On Polaris,
we use $D=K=256$ and 100 points per rank, so the three collectives operate on
$256$ integers (1\,KiB), $65{,}536$ doubles (512\,KiB), and 1 integer (4\,B) per
iteration, respectively. We evaluate $P=256$ (8 nodes) and $P=2048$ (64 nodes).
On Fugaku, we use $D=K=192$ with the same points-per-rank, yielding $192$ integers
(768\,B), $36{,}864$ doubles (288\,KiB), and 1 integer (4\,B) per iteration, and
evaluate $P=576$ (12 nodes) and $P=2304$ (48 nodes). Each configuration is executed
for 10 runs after 2 warmup runs; we report the per-run wall-clock time on the critical path
(maximum across ranks) and summarize run-to-run variability.

Figure~\ref{fig:kmeans-minmax-polaris-fugaku} shows that the iteration time is
strongly Allreduce-dominated, especially on Fugaku, and that substituting \tool
reduces the Allreduce portion directly, translating into end-to-end gains.
On Polaris, \tool reduces the total iteration time from 1.51\,s to 1.27\,s at
$P=256$ (1.19$\times$) and from 1.57\,s to 1.45\,s at $P=2048$ (1.08$\times$).
On Fugaku, where the collective fraction is larger, \tool reduces the iteration
time from 10.32\,s to 4.88\,s at $P=576$ (2.11$\times$) and from 11.20\,s to 5.04\,s
at $P=2304$ (2.22$\times$). These results confirm that \tool improves not only microbenchmarks but also the
runtime of a real iterative workload whose performance is bottlenecked by repeated
global reductions.

\vspace{-0.2cm}
\section{Related Work}
\label{sec:related}


Hierarchy-aware and node-aware collectives in MPI libraries aim to
exploit fast locality domains while limiting expensive cross-domain traffic, often
by aggregating locally and restricting the off-domain collective to a subset of
processes. Bienz \emph{et al.} analyze node-aware Allreduce schemes in which only
few ranks participate in the inter-node phase; while this reduces redundant
off-node messages, it can concentrate injection/progress responsibility and leave
many ranks idle, particularly for small reductions \cite{BienzNodeAware2019}. A related
modeling perspective is provided by Tr{\"a}ff’s k-lane formulation, which makes
explicit that multiple processes per node can act as concurrent ``lanes'' for
inter-node communication, and contrasts this capability with k-ported assumptions
for hierarchical networks \cite{traeff2020}. \tool adopts the same guiding
principle, avoiding a single fixed leader on the cross-domain path, by distributing
inter-batch communication across lanes and rotating root across
stages.

A further line of work shows that collective performance depends on
tunable algorithmic parameters and that robust performance requires selection
policies rather than a single fixed template. Faraj and Yuan demonstrate that
automatically generating and empirically tuning collective variants can improve
robustness across platforms and topologies \cite{faraj_2005}, while Wilkins
\emph{et al.} argue for generalized, parameterized collective families (including
variable-radix constructions) to span depth--fanout trade-offs on modern systems
\cite{exascale}. \tool aligns with this philosophy by exposing a small set of
hardware-relevant knobs (batch size and independent intra-batch radices) while
retaining a fixed, analyzable hierarchical schedule. Orthogonal approaches either
synthesize topology-specific collective schedules \cite{cai_2021} or reduce host
overhead via in-network reduction mechanisms such as SHArP and Flare
\cite{graham_2020,desensi_2021}; \tool instead focuses on a portable host-driven
design that can be deployed and tuned within standard MPI environments.

\bibliographystyle{IEEEtran}
\bibliography{bibliography}

\end{document}